\documentclass[conference,letterpaper]{IEEEtran}

\usepackage[utf8]{inputenc} 
\usepackage[T1]{fontenc}
\usepackage{url}
\usepackage{ifthen}
\usepackage{cite}
\usepackage[cmex10]{amsmath} 
\usepackage{amssymb}
\usepackage{graphicx}
\usepackage{multirow}
\usepackage{array}
\usepackage{makecell}
\usepackage{tikz}
\usepackage[ruled, vlined]{algorithm2e}
\usepackage{comment}
\usepackage[english]{babel}

\begin{document}
\title{A Modified Q-Learning Algorithm for Rate-Profiling of 
Polarization Adjusted Convolutional (PAC) Codes} 
	
\author{\IEEEauthorblockN{Samir Kumar Mishra, Digvijay Katyal and Sarvesha Anegundi Ganapathi}
\IEEEauthorblockA{\textit{Samsung Semiconductor India Research} \\
Bangalore, Karnataka, India \\
\{samir.mishra,  d.katyal, sarvesha.ag\}@samsung.com}
}
\maketitle

\begin{abstract}
In this paper, we propose a reinforcement learning based algorithm for rate-profile construction of Arikan's Polarization Adjusted Convolutional (PAC) codes. This method can be used for any blocklength, rate, list size under successive cancellation list (SCL) decoding and convolutional precoding polynomial. To the best of our knowledge, we present, for the first time, a set of new reward and update strategies which help the reinforcement learning agent discover much better rate-profiles than those present in existing literature.

Simulation results show that PAC codes constructed with the proposed algorithm perform better in terms of frame erasure rate (FER) compared to the PAC codes constructed with contemporary rate profiling designs for various list lengths. Further, by using a (64, 32) PAC code as an example, it is shown that the choice of convolutional precoding polynomial can have a significant impact on rate-profile construction of PAC codes.
\end{abstract}

\section{Introduction}
\IEEEPARstart{P}{olar} codes introduced by Arikan \cite{arikan2009channel} are the first provably capacity achieving codes for the class of binary input memoryless symmetric (BMS) channels with low encoding and decoding complexity of order $ O(N\log_2{N})$ for a code of blocklength $N$. Polar code is based on a phenomenon of channel polarization where a communication channel is transformed into polarized sub-channels: either completely noisy or noiseless. Information bits are transmitted over a set of noiseless sub-channels, while fixed or \textit{frozen} bits are sent over the noisy ones. Polar codes are already being used in 5G New radio (NR) \cite{3gpp.38.212} for encoding and decoding of control information.

\begin{figure}
\centering
\includegraphics[width=3.4in]{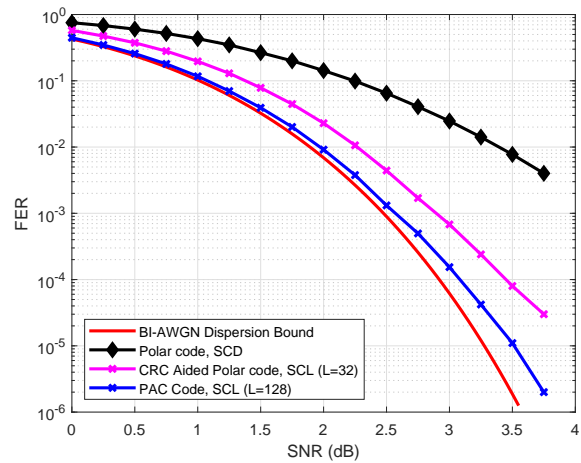}
\caption{FER Performance of (128, 64) polar code variants}
\label{fig1}
\end{figure}

Polar codes achieve channel capacity asymptotically as the blocklength $N$ of the code approaches infinity. However, for short blocklengths, the performance of polar codes is not good enough. Figure \ref{fig1} shows the performance of polar code and its variants for a blocklength $N=128$ and rate $R=0.5$ for a binary input additive white gaussian noise (BI-AWGN) channel. Figure \ref{fig1} also shows the BI-AWGN dispersion bound which is the minimum probability of error $\epsilon^{*}(N, R)$ that can be achieved on a BI-AWGN channel by using a code of blocklength $N$ and rate $R$ under maximum likelihood (ML) decoding. It can be observed clearly that there is a big gap between polar code with successive cancellation decoding (SCD) and the dispersion bound. This poor performance can be partly attributed to poor distance properties of polar codes and also the sub-optimality of SCD as compared to ML decoding \cite{tal2015list}.

Since Arikan's ground breaking work, there have been a lot of efforts to enhance the performance of polar code for short blocklengths, a survey of which can be found in \cite{cocskun2018efficient}. Specifically, CRC-Aided Polar codes under succesive cancellation list (SCL) decoding \cite{tal2015list} improve the performance quite a lot. Figure \ref{fig1} shows the FER performance of a (128, 72) polar code combined with a (72, 64) cyclic code which acts as CRC under SCL decoding with a list size $L=32$. This approach has been adopted to 5G NR standard and has remained the state of the art ever since.

In the Shannon Lecture at the International Symposium on Information Theory (ISIT) 2019, Arikan presented \textit{polarization adjusted convolutional (PAC) codes}, which are significant improvement over the state-of-the-art polar codes \cite{arikan2019sequential}. Under sequential decoding, the FER performance of PAC codes is just 0.25 dB away from the BI-AWGN dispersion bound approximation at a target FER of $10^{-5}$. Further, it was observed in \cite{yao2020list} and \cite{rowshan2020polarization} that nearly same FER performance can as well be obtained by list decoding as shown in Figure \ref{fig1}.

\subsection{Brief overview of PAC code}
\begin{figure}
\centering
\includegraphics[width=3.4in, height=1.7in]{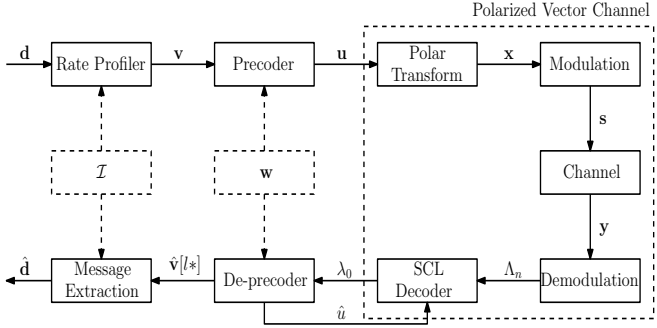}
\caption{Coding scheme of PAC code}
\label{fig2}
\end{figure}

The coding scheme of a PAC code is shown in Figure \ref{fig2}. In Figure \ref{fig2}, the solid blocks refer to actual blocks used in the communication system. The dotted blocks refer to the information provided to these blocks. A PAC code can be denoted as $PAC(N, K, \mathcal{I}, \mathbf{w})$. Here, $K$ is the number of information bits. $N$ is the length of the codeword which is mostly a power of 2. $\mathcal{I} \subseteq \{ 0, 1, ..., (N-1)\}$ is the set of information bit indices. $\mathbf{w}$ is a precoding vector of length $p$ containing 0s and 1s. $R=K/N$ is the rate of the code. Using the polar code terminology, $\mathcal{F} = \mathcal{I}^{c}$ is the set of \textit{frozen} indices, where no information is transmitted. These indices are filled with zeros.

\subsubsection{PAC Encoding scheme}
The first step of encoding PAC codes is rate-profiling, a term coined by Arikan\cite{arikan2019sequential}. A rate-profiler maps the vector of information bits denoted by $\mathbf{d} = [d_0, d_1, ..., d_{K-1}]$ to a vector of bits $\mathbf{v} = [v_0, v_1, ..., v_{N-1}]$ according to $\mathcal{I}$. In other words, the $K$ information bits in $\mathbf{d}$ are mapped to positions in $\mathbf{v}$ indicated by $\mathcal{I}$. The rest $(N-K)$ positions in $\mathbf{v}$ are filled with zeros. The selection of $K$ indices out of $N$ possible indices is called rate-profile construction. Two well known rate-profiling schemes are polar and Reed-Muller (RM) rate profiles.

After rate-profiling, the vector $\mathbf{v}$ is transformed into a vector $\mathbf{u}$ using a convolutional precoding polynomial $\mathbf{w}$ of length $p$. In other words, each bit in $\mathbf{v}$ is replaced by a linear combination of itself and $p-1$ bits that precede it. This linear combination is decided by $\mathbf{w}$.

The final step of encoding is to pass the precoded vector $\mathbf{u}$ through a Polar Transform $\mathbf{P}_n$ to output encoded bit vector $\mathbf{x} = \mathbf{u}\mathbf{P}_{n} = \mathbf{u}\mathbf{P}^{\bigotimes n}$. Here, $\mathbf{P}_n$ is the $n^{th}$ Kronecker power of the basic Polar Transform $\mathbf{P}= \begin{bmatrix}
			1 & 0 \\
			1 & 1
			\end{bmatrix}$ proposed by Arikan in \cite{arikan2009channel}.
			
In the absence of precoding, $\mathbf{w} = [1]$ and PAC code falls back to polar code.

\subsubsection{PAC Decoding scheme}:
The decoding of PAC code can be done either by sequential decoding\cite{arikan2019sequential} or by list decoding as in \cite{yao2020list}, \cite{rowshan2020polarization}, \cite{mishra2020selectively}.

\subsection{Our Contributions}
It was observed by Arikan in \cite{arikan2019sequential} that the performance of PAC code is more sensitive to the choice of rate-profiling scheme $\mathcal{A}$ as compared to the precoding polynomial $\mathbf{w}$. Further, it was also observed in \cite{yao2020list}, that PAC codes perform better than polar codes owing to improved minimum distance properties.

In this paper, we present a reinforcement learning (RL) algorithm for rate-profile construction of PAC code that can be used for any blocklength, rate, list size (under SCL decoding) and precoder constraints. The algorithm tries to construct a rate-profile $\mathcal{I}$ for a $(N, K, \mathcal{I}, \mathbf{w})$ PAC code by mapping polar code construction to a maze traversing game \cite{9348092}. Contrary to \cite{9348092}, where the authors use a SARSA($\lambda$) algorithm, we propose to use the modified version of a much simpler reinforcement learning algorithm, that is, Q-Learning to solve the maze traversing game. To the best of our knowledge, we present, for the first time, a set of new reward and update strategies which help the RL agent discover much better rate-profiles than those present in existing literature.

In order to validate our claims, we compare the PAC code constructed with the proposed algorithm with the PAC codes available in existing literature. Simulation results show that PAC codes with the proposed rate-profile construction perform better in terms of frame erasure rate (FER) compared to the PAC codes with rate profiling designs in existing literature. 

In this paper, vectors are denoted by boldface lowercase letters $\mathbf{a}$. An element in a vector $\mathbf{a}$ at index $i$ is denoted by $a_i$. A set is denoted by caligraphic font $\mathcal{A}$, its cardinality by $|\mathcal{A}|$ and its complement by $\mathcal{A}^{c}$.

The rest of the paper is organized as follows. In Section II, we present our proposed heuristic algorithm. Section III discusses simulation results for some PAC codes constructed by using the proposed algorithm and compare it with schemes in existing literature. Finally, in Section IV, we conclude by mentioning some open problems.

\section{VIEWING POLAR CODE CONSTRUCTION AS A GAME}
The construction of a $(N,K)$ polar code is the selection of $K$ non-frozen bit positions out of $N$ bit positions. This selection procedure can be viewed as a maze traversing game in the reinforcement learning setup, where an agent tries to find the optimal path for the given environment, rewards and set of rules. 

\begin{algorithm}\label{algo_qlearning}
    \SetKwInOut{Input}{Input}
    \SetKwInOut{Output}{Output}
    
    \Input{$N$, $K$, $\mathbf{w}$, $p$}
    \Output{$Q$}
    States $\mathcal{S} = \{(0,1),(0,1),(0,2),\dots,(N-K,K)\}$\\
    Actions $\mathcal{A} = \{0,1\}$\\
    NextState: $\mathcal{S}\times\mathcal{A}\longrightarrow\mathcal{S}$\\
    $Q: \mathcal{S}\times\mathcal{A}\longrightarrow\mathbb{R}$\\
    Episodes $E\gets\mathbb{N}$\\
     Discounting factor $\gamma\gets 1$\\
    Learning rate $\alpha\in[0,1]$\\
    Exploration rate $\epsilon\in[0,1]$\\
    \For{$e \gets 0$ \KwTo $(E-1)$}{
    $s \gets$ initial state\\
    $\mathcal{I}_{init}, \mathcal{N} \gets \text{RMScore}(N, K)$\\
    \For{$k \gets 0$ \KwTo $(N-1)$}{
    $\mathcal{I}_{init}$, $\mathcal{N}$, $a \gets  \text{Action}(Q, \mathcal{I}_{init}, \mathcal{N}, K, k, \epsilon)$\\
    $a_k \gets a$\\
    $s' \gets  \text{NextState}(s, a)$\\
    \If{$k=0$}{
    Transmit all zero codeword through the channel.
    }
    \If{$a=0$}{
    Decode the $k^{th}$ bit as if it is a frozen bit.
    Update the PM list and the survival paths.
    }\Else{
    Decode the $k^{th}$ bit as if it is a non-frozen bit.
    Update the PM list and the survival paths.
    Check if all-zero codeword survives in the list.\\
    \If{all zero codeword dropped}{
    $r \gets  -2x$\\
    $Q \gets  \text{Update}(N, K, Q, s, s', a, r, \alpha, \gamma)$\\ 
    $F \gets 1$\\
    \textbf{break}
    }\Else{
    $F \gets 0$\\
    }
    }
    $s \gets s'$\\
    }
    \If{$F=0$}{
    $i \gets$ Get the index of all zero codeword in the PM list\\
    $\mathbf{c} \gets$ Get the first codeword in the PM list\\
    $s \gets$ initial state\\
    $f \gets 0$\\
    \For{$k \gets 1$ \KwTo $N$}{
    $a \gets a_k$\\
    $s' \gets  \text{NextState}(s, a)$\\
    $r, f \gets \text{Reward}(\mathbf{c}, k, i, f)$\\
    $Q \gets  \text{Update}(N, K, Q, s, s', a, r, \alpha, \gamma)$\\
    $s \gets s'$\\
    }
    }
    }
    
    \Return $Q$

    \caption{Proposed modified Q-learning algorithm}
\end{algorithm}

\subsection{Environment}
Each $(N,K)$ polar code construction problem is viewed as a maze with $N-K+1$ rows and $K+1$ columns. This maze is equivalent to environment in reinforcement learning problem. Cells of maze define the states $s=(row,col)$ of the environment. At any time, the RL agent can be in one of the possible state $s$, where, $s \in \mathcal{S}, |\mathcal{S}| = (N-K+1)\times(K+1) $. State $s = (0,0)$ is defined as the initial state and state $s=(N-K,K)$ is defined as the terminal state. At each state $s$ the agent can take one of two possible actions $a$ where $a \in \mathcal{A}, |\mathcal{A}| = 2$ i.e., ``down'' action and ``right'' action.  For each episode, the agent is required to start from the initial sate and ends up at terminal state by taking in total $N$ steps.  

Our proposed rate-profile construction algorithm is presented in \textbf{Algorithm} \ref{algo_qlearning}. The inputs to the algorithm are $N$, $K$, $\mathbf{w}$ and $p$. The rate-profile for the intended PAC code can be derived as the best path in the maze discovered by the RL agent and can be computed from the output  $Q$ of the algorithm. We describe the details of this algorithm in the forthcoming paragraphs.

\subsection{Action Strategy}
As already mentioned, the aim of rate-profile construction is to select $K$ indices out of $N$ possible indices. Also at each step, the agent is required to select one of the two possible actions i.e., ``down'' $(a=0)$ and ``right'' $(a=1)$ action. In particular, if at $k^{th}$ step, the agent selects down action, then $k^{th}$ bit  corresponds to frozen bit position and if the agent selects right action then the $k^{th}$ bit corresponds to non-frozen/information bit position.  Our proposed action generation strategy as described in \textbf{Algorithm} \ref{algo_action} can divided into two main phases as described below.

We know that there is a RM score corresponding to each of the $N$ indices. In the first phase, the RM score generation strategy described in \textbf{Algorithm} \ref{algo_rmscore} sorts the RM scores of $N$ indices in an ascending order. We call the RM score at the $(N-K)^{th}$ index of the sorted set as \textit{boundary RM score}. Now, we can divide the set of N indices into three subsets. First, for the set of indices whose RM score is less than the boundary RM score, the agent will take the down action, i.e., these indices are allocated to the \textit{frozen set} $\mathcal{F}$ or $\mathcal{I}^{c}$. Second, for the set of indices whose RM score is greater than the boundary RM score, agent will take the right action i.e, these indices are allocated to the set of \textit{information bit indices} $\mathcal{I}$. We call this set as the \textit{initial rate profile} and denote it by $\mathcal{I}_{init}$. It is to be noted that $|\mathcal{I}_{init}| \le K$. 

\begin{algorithm}\label{algo_rmscore}
    
    \SetKwFunction{FMain}{RMScore}
    \SetKwProg{Fn}{subroutine}{:}{}
    \Fn{\FMain{$N$, $K$}}{
    $\mathbf{t} \gets \mathbf{0}$ \\
    \For{$j \gets 0$ \KwTo $(N-1)$}{
      $t_j \gets w(j-1)$ \tcp*[f]{Calculate RM score} \\
      }
     $\mathbf{u}$ = sort($\mathbf{t}$) \tcp*[f]{Sort in ascending order}\\
     $t_b \gets \mathbf{u}_{N-K}$ \tcp*[f]{Boundary RM score}\\
     $\mathcal{I}_{init} \gets  \emptyset$  \tcp*[f]{Initial rate profile}\\
      $\mathcal{N} \gets  \emptyset$ \tcp*[f]{Set of indices  with $t_i = t_b$}\\
     \For{$i \gets 0$ \KwTo $(N-1)$}{
      \If{$t_i > t_b$}{
        $\mathcal{I}_{init} \gets  \mathcal{I}_{init} \cup \{i\}$ \\
      }
      \If{$t_i = t_b$}{
        $\mathcal{N} \gets  \mathcal{N} \cup \{i\}$ \\
      }
     }
     
    \Return $\mathcal{I}_{init}, \mathcal{N}$
    }

    \caption{RM Score generation algorithm}
\end{algorithm}

An obvious simplification happens for some PAC codes when $|\mathcal{I}_{init}| = K$. The code is constructed by the first phase of the algorithm itself. A typical example of this is the $(128, 64)$ PAC code. In this case, our rate-profile construction algorithm falls back to the famous RM rate-profile as described by Arikan in \cite{arikan2019sequential}.

Third, we have the set of indices whose RM score is equal to the boundary RM score. We denote this set by $\mathcal{N}$. In the second phase, the action generation algorithn selects the remaining $(K - |\mathcal{I}_{init}|)$ indices from available $|\mathcal{N}|$ indices in order to create a $(N, K)$ code. For these set of indices, the agent selects the action $a$ from the current state $s \in \mathcal{S}$ using policy derived from $Q$ (e.g. $\epsilon$-greedy).

\begin{algorithm}\label{algo_action}
    \SetKwFunction{FMain}{Action}
    \SetKwProg{Fn}{subroutine}{:}{}
    \Fn{\FMain{$Q$, $\mathcal{I}_{init}$, $\mathcal{N}$, $K$, $k$, $\epsilon$}}{
    \If{$k \in \mathcal{N}$}{
    \If{$|\mathcal{I}_{init}| = K$}{
    $a \gets 0$\\
    }\ElseIf{$|\mathcal{I}_{init}|+|\mathcal{N}|= K$}{
    $a \gets 1$\\
    }\Else{
    $a \gets \epsilon \text{-greedy}(Q, \epsilon)$\\
    \If{$a=0$}{
    $\mathcal{N} \gets \mathcal{N} \setminus \{k\}$ \\
    }\Else{
    $\mathcal{I}_{init} \gets  \mathcal{I}_{init} \cup \{k\}$ \\
    }
    }
    }\ElseIf{$k \in \mathcal{I}_{init}$}{
    $a \gets 1$
    }\Else{
    $a \gets 0$ 
    }
    \Return $\mathcal{I}_{init}$, $\mathcal{N}$, $a$
    }

    \caption{Proposed action generation algorithm}
\end{algorithm}

\subsection{Update}
Value function of state action pairs are updated when either all zero codeword is dropped from the list or the episode comes to an end. In the first case, the value function of the current state action pair is updated and on the other hand in later case, the value function of all the state action pairs taken during that episode are updated. The proposed update strategy is described in \textbf{Algorithm} \ref{algo_update}.

\begin{algorithm}\label{algo_update}
   \SetKwFunction{FMain}{Update}
    \SetKwProg{Fn}{subroutine}{:}{}
    \Fn{\FMain{$N$, $K$, $Q$, $s$, $s'$, $a$, $r$, $\alpha$, $\gamma$}}{
    \If{$s_{0}' = N-K$}{
    $Q(s,a) \gets Q(s,a)+\alpha(r+\gamma Q(s',1) - Q(s,a))$\\
    }\ElseIf{$s_{1}' = K$}{
    $Q(s,a) \gets Q(s,a)+\alpha(r+\gamma Q(s',0) - Q(s,a))$\\
    }\Else{
    $Q(s,a) \gets Q(s,a)+\alpha(r+\gamma \max_{a'} Q(s',a') - Q(s,a))$\\
    }
    \Return $Q$
    }
    \caption{Proposed update strategy algorithm}
\end{algorithm}

\subsection{Reward Strategy}
If all zero codeword survives in the list through out the episode i.e., $F=0$, value function $Q(s,a)$ of all the $N$ state action pairs taken during that episode are updated according to the update strategy described in \textbf{Algorithm} \ref{algo_update}, using the reward generating strategy described in \textbf{Algorithm} \ref{algo_reward}. 

Reward for the $k^{th}$ state action pair depends on two factors. Firstly, it depends on the $k^{th}$ output bit $c_k, c_k \in \{0,1\}$. Secondly, it depends on the index in the list where the all zero codeword is present if not dropped from the list. We call this index as \textit{all zero codeword index} and denote it by  $i, i \in [1, L]$ in the list. A positive  reward of $x$, defined as ``base'' reward is given, if the $k^{th}$ bit is correctly decoded, otherwise a negative reward of $-x$ is given. Based on the all zero codeword index, reward is drop by amount $z(i-1)$, where $z$ is known as the ``step'' reward  and $i$ is the index of all zero codeword in the list at the end of the episode. Also, if the $k^{th}$ bit position corresponds to first bit error position in output $\mathbf{c}$, reward is further drop by amount $x$.

In case, if all zero codeword dropped from the list i.e., $F=1$, value function $Q(s,a)$ of current state action pair will updated according to the update strategy described in \textbf{Algorithm} \ref{algo_update} using the reward of $-2x$. 

It must be noted here that the intention of the above mentioned reward strategy is to make the RL agent learn rate-profiles which help to keep the path with all zero codeword as high in the list as possible so that it becomes the best path.

\begin{algorithm}\label{algo_reward}
   \SetKwFunction{FMain}{Reward}
    \SetKwProg{Fn}{subroutine}{:}{}
    \Fn{\FMain{$\mathbf{c}$, $k$, $i$, $f$}}{
    \If{$c_k=0$}{
    $r \gets x - z(i-1)$
    }\Else{
    $r \gets -x - z(i-1)$\\
    \If{$f=0$}{
    $r \gets r - x$\\
    $f \gets 1$\\
    }
    }
    \Return $r, f$
    }
    \caption{Proposed reward generation algorithm}
\end{algorithm}

\section{Numerical results and Discussion}
\begin{figure}
\centering
\includegraphics[width=3.4in]{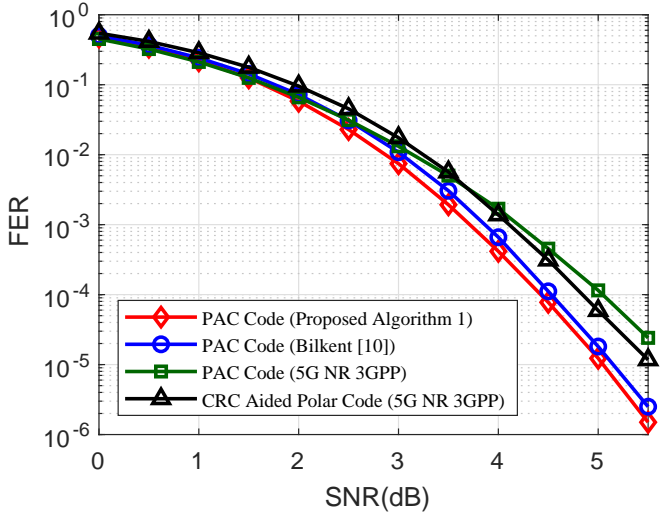}
\caption{FER performance of $(64, 32)$ CRC-Aided polar code and PAC code variants with $\mathbf{w}=[1,0,1,1,0,1,1]$ under SCL decoding with $L=8$.}
\label{fig3}
\end{figure}

For simulation, we first consider a (64, 32) PAC code transmitted over a BI-AWGN channel for which the rate-profile $\mathcal{I}$ was constructed using the proposed method. The convolutional precoding polynomial used is $\mathbf{w} = [1,0,1,1,0,1,1]$. In figure \ref{fig3} we compare the FER performance of CRC-Aided polar code and PAC code variants in existing literature with that of the PAC code constructed using our proposed Algorithm \ref{algo_qlearning}. For each code, SCL decoder with list size $L = 8$ was used. The algorithm proposed in \cite{moradi2021monte} performs much better than the current state-of-the-art 5G NR rate profile. Further, it can be easily observed that the PAC code constructed with our proposed Algorithm \ref{algo_qlearning} outperforms all contemporary PAC code constructions at all values of SNR. Further, we compare proposed PAC code with the current state-of-the-art 8-bit CRC-Aided (64, 32) Polar code used in 3GPP 5G NR which contains 24 information bits and 8 CRC-bits. Specifically, at a target FER of $10^{-5}$, the performance of proposed PAC code is around 0.5 dB better than the CRC-Aided Polar code.

\begin{figure}
\centering
\includegraphics[width=3.4in]{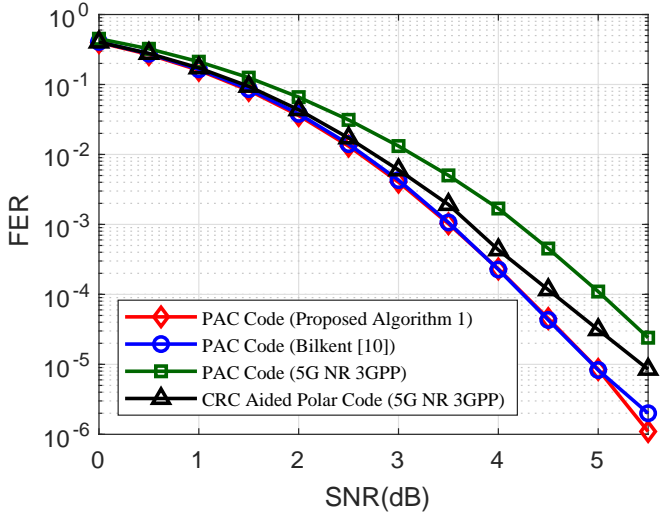}
\caption{FER performance of $(64, 32)$ polar code, CRC-Aided polar code and PAC code variants with $\mathbf{w}=[1,0,1,1,0,1,1]$ under SCL decoding with $L=32$.}
\label{fig4}
\end{figure}

Figure \ref{fig4} shows (64, 32) PAC code variants with same parameters as those shown in Figure \ref{fig3}. However, we have used SCL decoder with list size $L = 32$. It can be observed that both proposed PAC code and algorithm proposed in \cite{moradi2021monte} perform nearly same. Further, we compare proposed PAC code with the current state-of-the-art 8-bit CRC-Aided (64, 32) Polar code used in 3GPP 5G NR which contains 24 information bits and 8 CRC-bits.  Specifically, at a target FER of $10^{-5}$, the performance of proposed PAC code is around 0.5 dB better than the CRC-Aided Polar code.

\begin{figure}
\centering
\includegraphics[width=3.4in]{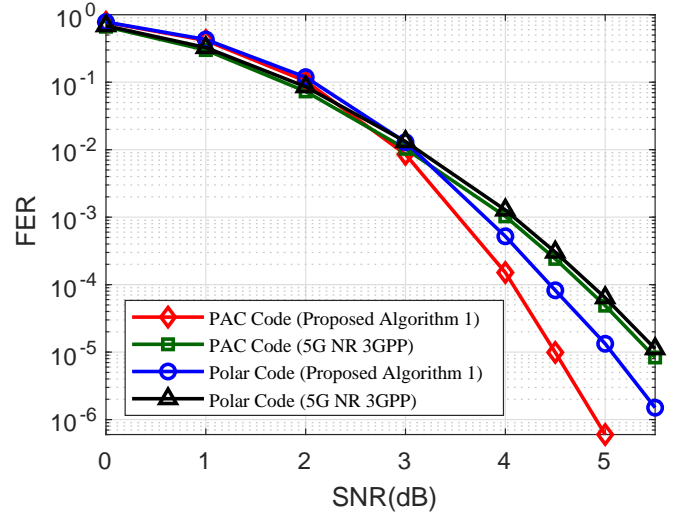}
\caption{FER performance of $(128, 72)$ polar code and PAC code variants with $\mathbf{w}=[1,0,1,1,0,1,1]$ under SCL decoding with $L=8$.}
\label{fig5}
\end{figure}

In figure \ref{fig5}, we consider a (128, 72) PAC code transmitted over a BI-AWGN channel for which the rate-profile $\mathcal{I}$ was constructed using Algorithm \ref{algo_qlearning}. The convolutional precoding polynomial used is $\mathbf{w} = [1,0,1,1,0,1,1]$. It can be seen that both polar and PAC code constructed using algorithm \ref{algo_qlearning} performs much better than current state-of-the-art 3GPP 5G NR \cite{3gpp.38.212} rate profile. Specifically, at a target FER of $10^{-5}$, the performance of proposed polar and PAC code is around 0.5 dB and 0.9 dB better than the  3GPP 5G NR rate profile.

\begin{figure}
\centering
\includegraphics[width=3.4in]{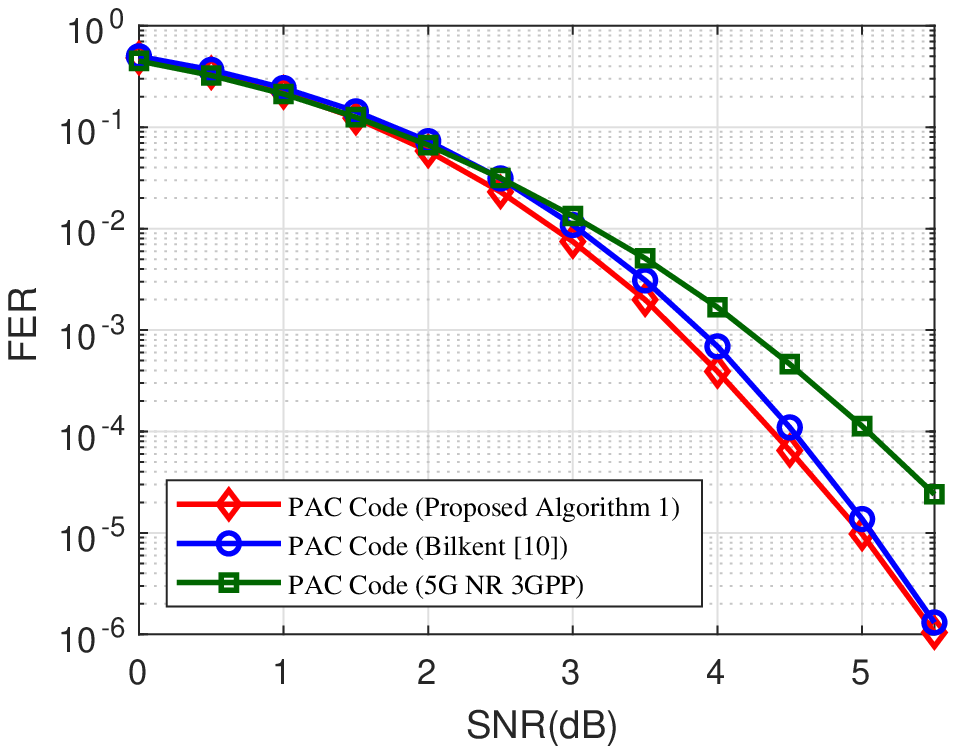}
\caption{FER performance of $(64, 32)$ polar code, CRC-Aided polar code and PAC code variants with $\mathbf{w}=[1,1,0,1,0,0,0,1,0,0,1]$ under SCL decoding with $L=8$.}
\label{fig6}
\end{figure} 

\begin{table*}
\caption{Rate Profiles}
  \begin{center}
  \renewcommand{\arraystretch}{2}
    \begin{tabular}{|c||c||c||c|}
    \hline
    $(N,K)$ & \textbf{Precoder} $\mathbf{w}$ \textbf{(binary)}& \textbf{List Size} $L$ & \textbf{Rate Profiles} $\mathcal{I}$ \textbf{(hexadecimal)}\\
    \hline
    \hline
    \multirow{3}{*}{$(64,32)$}
    & 1 & 8 & 01050377051F7F7F\\
    \cline{2-4}
    & 1011011 & 8 & 0015115F175717FF\\
    \cline{2-4}
    & 1011011 & 32 & 01070737057F177F\\
    \hline
    \hline
    \multirow{2}{*}{$(128,72)$}
    & 1 & 8 & 0001115701173F7F053F177F17FF7FFF \\
    \cline{2-4}
    & 1011011 & 8 & 0011011711371FFF0177577F177F7FFF \\
    \hline
    \hline
    \multirow{2}{*}{$(256,128)$}
     & 1011011 & 8 &  000100010001011F0001113F073737FF0105157F055F5F7F157F5FFF7FFFFFFF\\
    \cline{2-4}
    & 1011011 & 32 &  000100010001011F0001113F073737FF0105157F055F5F7F157F5FFF7FFFFFFF\\
    \hline
    
    \end{tabular}
  \end{center}
  \label{table1}
\end{table*}

Figure \ref{fig6} shows (64, 32) PAC code variants transmitted over a BI-AWGN channel, but using a convolution precoding polynomial given by $\mathbf{w}=[1,1,0,1,0,0,0,1,0,0,1]$. For each code, SCL decoder with list size $L= 8$ is used. In this case, PAC code constructed with our proposed Algorithm \ref{algo_qlearning} outperforms the PAC code constructed with algorithm in \cite{moradi2021monte} at all values of SNR.

\begin{figure}
\centering
\includegraphics[width=3.4in]{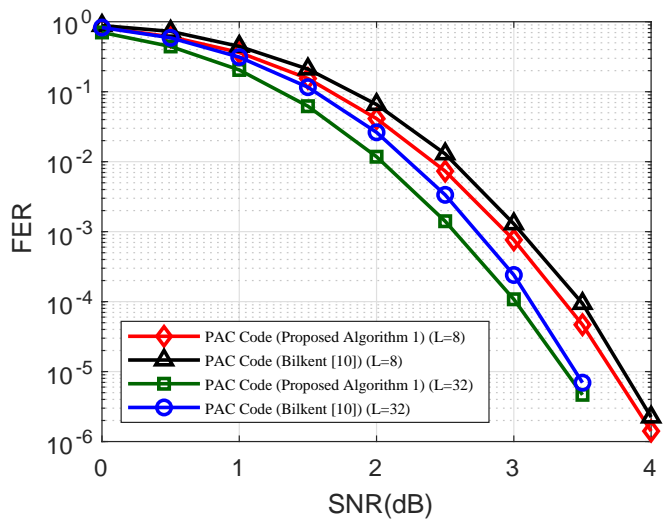}
\caption{FER performance of $(256, 128)$ polar code, CRC-Aided polar code and PAC code variants with $\mathbf{w}=[1,1,0,1,0,0,0,1,0,0,1]$ under SCL decoding with $L=8$ and $L=32$.}
\label{fig7}
\end{figure}

Figure \ref{fig7} shows the comparison of FER performance of $(256, 128)$ PAC code constructed using Algorithm \ref{algo_qlearning} versus the PAC code constructed using the algorithm mentioned in \cite{moradi2021monte}. The convolutional precoding polynomial used is $\mathbf{w} = [1,1,0,1,0,0,0,1,0,0,1]$. The improved performance of PAC code constructed with our proposed algorithm is evident in this case as well.

Table \ref{table1} provides the rate-profiles that were obtained by Algorithm \ref{algo_qlearning} and used to generate the simulation results just described.


\section{Conclusion and Future Work}

In this paper, we presented a modified Q-Learning algorithm for rate-profile construction of Arikan's PAC codes. We demonstrated that the PAC codes constructed by using the proposed algorithm perform better in terms of FER compared to the PAC codes constructed with rate-profile designs in existing literature. It was shown how the choice of the convolutional precoding polynomial can affect the performance of PAC code. 

There are many open problems to ponder. First, we note that the proposed RL algorithm  may terminate at a locally optimal solution. Sometimes, this might lead to poor construction of PAC codes. Is it possible to create a better rate-profile construction algorithm?  Second, for a given code rate and blocklength, does there exist a convolutional precoding polynomial for which the FER at a given SNR is minimum? Third, the rate-profile construction algorithms in current literature including the proposed algorithm in current paper do not consider rate-matching \cite{3gpp.38.212} schemes which are needed for practical application of PAC codes. We consider these problems for future work.

\bibliographystyle{IEEEtran}
\bibliography{ref}

\end{document}